\documentclass{article}
\usepackage[utf8]{inputenc}
\usepackage[margin=1in]{geometry}
\usepackage[dvipsnames]{xcolor}
\usepackage{amsmath}
\usepackage{physics}
\usepackage{graphicx}
\usepackage{amssymb}
\usepackage{enumitem}
\usepackage{mathtools}
\usepackage{authblk}
\usepackage{siunitx}
\usepackage{hyperref}
\hypersetup{colorlinks=true, citecolor=red, urlcolor=blue, linkcolor=blue}
\usepackage{caption}

\newcommand{\disAv}[1]{\mathbb{E}\left[  {#1} \right] }
\newcommand{\G}{G}
\newcommand{\SFF}{\textrm{SFF}}

\newcommand{\trp}{\textrm{TRP}}
\newcommand{\rt}{\tilde R}

\title{\bf Reappearance of Thermalization Dynamics in the Late-Time Spectral Form Factor}
\author[1]{Michael Winer}
\author[2]{Brian Swingle}
\affil[1]{Joint Quantum Institute, Department of Physics, University of Maryland, College Park, Maryland 20742, USA}
\affil[2]{Department of Physics, Brandeis University, Waltham, Massachusetts 02453, USA}

\begin{document}
\maketitle
\begin{abstract}
    The spectral form factor (SFF) is an important diagnostic of energy level repulsion in random matrix theory (RMT) and quantum chaos. The short-time behavior of the SFF as it approaches the RMT result acts as a diagnostic of the ergodicity of the system as it approaches the thermal state. In this work we observe that for systems without time-reversal symmetry, there is a second break from the RMT result at late time around the Heisenberg time. Long after thermalization has taken hold, and after the SFF has agreed with the RMT result to high precision for a time of order the Heisenberg time, the SFF of a large system will briefly deviate from the RMT behavior in a way exactly determined by its early time thermalization properties. The conceptual reason for this second deviation is the Riemann-Siegel lookalike formula, a resummed expression for the spectral determinant relating late time behavior to early time spectral statistics. We use the lookalike formula to derive a precise expression for the late time SFF for semi-classical quantum chaotic systems, and then confirm our results numerically for more general systems.
\end{abstract}

\section{Introduction}

The problem of quantum thermalization is one of the key questions in many-body physics. One of the most important diagnostics of thermalization is spectral statistics \cite{haake2010quantum,PhysRevLett.52.1,mehta2004random}---the statistical properties of the energy eigenvalues of the Hamiltonian. There is a great deal of evidence that the spectra of ensembles of Hamiltonians of chaotic systems share certain universal features with the spectra of ensembles of Gaussian random matrices (RMT), with examples ranging from condensed matter theory~\cite{bohigas1984chaotic,dubertrand2016spectral,PhysRevLett.121.264101,Chan:2017kzq,PhysRevX.8.021062}, to nuclear theory~\cite{doi:10.1063/1.1703775,wigner1959group}, to field theory~\cite{PhysRevLett.126.121602,Delacretaz:2022ojg}, to holography~\cite{Cotler2017,Saad:2018bqo}. This phenomenon is known as random matrix universality. Yet physical systems are not literally described by Gaussian random Hamiltonians, so it is crucial to understand the physical origins and limits of random matrix universality. 

Spectral statistics can be usefully diagnosed using a powerful tool called the Spectral Form Factor (SFF). This quantity indicates whether energy levels repel as they do in random matrices~\cite{bohigas1984chaotic}, have independent Poissonian statistics~\cite{berry1977level}, or have some more exotic behavior ~\cite{KunzShapiro,PhysRevLett.125.250601,RichardGlass,Winer_Glass}. 
The SFF is defined as
\begin{equation}
    \SFF(T,f)=\disAv{\tr(f(H)e^{iHT})\tr(f(H)e^{-iHT})}=\disAv{\sum_{nm}f(E_m)f(E_n)e^{i(E_m-E_n)T}},
    \label{eq:sffDef}
\end{equation}
where $\disAv{\cdots}$ denotes a disorder average over an ensemble of Hamiltonians and $f$ is a filter function that can be used to address specific parts of the spectrum. 
\begin{figure}
    \centering
    \includegraphics[scale=0.9]{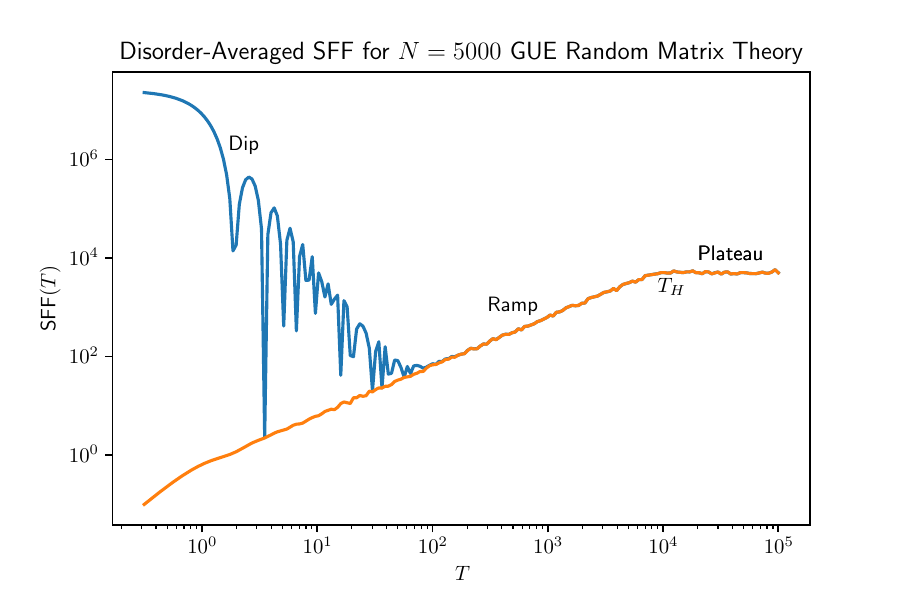}
    \caption{Blue: A log-log plot of the SFF of a GUE random matrix of size $5000$ with $f=1$. The oscillating `dip' region comes from the sharp boundaries of Wigner's semicircle law. The long linear `ramp' turns sharply into a `plateau' at time $T_H=2\pi \bar \rho$ where $\bar{\rho}$ is the average density of states. Orange: The connected SFF, as defined in equation \eqref{eq:SFFcon}.}
    \label{fig:sffPlot}
\end{figure}
Figure \ref{fig:sffPlot} shows the $f=1$ SFF of a random matrix ensemble of $5000 \times 5000$ complex Hermitian matrices. This ensembles corresponds to generic Hermitian operators without time-reversal or other symmetries, and it is referred to as the Gaussian unitary ensemble (GUE). The early time region known as the `dip' depends on the precise shape of the density of states and on $f$ and is thus less universal; by contrast, the ramp and plateau are quite generic. The presence or absence of the ramp in the SFF at sufficiently late time is the prime diagnostic of random matrix universality.

In addition to detecting level repulsion, a more careful accounting \cite{PhysRevLett.121.264101,WinerHydro,bipartite,Friedman_2019,moudgalya2020spectral} shows that the behavior of the spectral form factor at time $T$ contains information about thermalization at time $T$. A pure RMT ramp indicates complete thermalization, whereas many physical systems exhibit a time-dependent enhancement of the ramp that corresponds to incomplete thermalization. Quantitatively, the SFF at time $T$ is multiplied by a factor called the Total Return Probability (TRP) which is large when the system remembers its initial state but goes to one when the system thermalizes. The time-scale for full thermalization can be long~\cite{Gharibyan_2018,Friedman_2019,PhysRevLett.121.060601,WinerHydro}, even depending on the system size in some way, but one might expect that once thermalization has occurred, as diagonosed by the TRP, then the SFF will follow the pure RMT result for all subsequent times.

This expectation turns out to be wrong: the early time enhancement must come with a cost. As a matter of Fourier analysis, one can show that for any system with level repulsion the total SFF integrated over all times $T$ is a function only of the density of states (see Section \ref{sec:SFFInt} for more details). This means that the enhancement at short times must be `paid back' later via a suppression in the SFF relative to the RMT result. In this work we investigate systems without time reversal symmetry (GUE-type) and conjecture that the early time enhancement is precisely matched by a late time suppression around the Heisenberg time ($2\pi$ times the inverse level spacing). This bizarre resurgence of early-time dynamics for a brief window had gone unnoticed because the corrections are numerically smaller than the plateau value of the SFF. But we provide a precise formula for the corrections which can be confirmed numerically, as in Figure~\ref{fig:numericalConfirmation}. Although the formula itself is more general, we also show how to derive it analytically in the special case of systems admitting a semi-classical description in terms of periodic orbits. This approach has a long history, including~\cite{Gutzwiller:1971fy,Berry1981,MVBerry_1990,10.2307/52022,Keating_2007,PhysRevLett.89.206801,2009NJPh...11j3025M,2005PhRvE..72d6207M,Braun_2012}, and we find it useful to start from \cite{Keating_2007,2009NJPh...11j3025M,Braun_2012}.

The derivation uses a remarkable formula known as the Riemann-Siegel lookalike~\cite{10.2307/52059,MVBerry_1990,10.2307/52022,bogomolny}. Inspired by properties to the Riemann zeta function, this lookalike is a resummation formula which provides a quantitative link between early- and late-time behaviors of spectral statistics. With this formula in hand one can derive a new expression for the behavior of the SFF in systems without time reversal, one that works at all times. 

To state our main result, we need the density of states
\begin{equation}
    \rho(E)=\sum_n \delta(E-E_n),
\end{equation}
its average
\begin{equation}
    \bar{\rho}(E) =\disAv{\rho(E)} 
\end{equation}
and a kernel $K(T,E)$ in terms of which the SFF is
\begin{equation}
    \SFF(T,f)= \int dE f(E)^2 K(T,E) + \left| \int dE f(E) \bar{\rho}(E) e^{-i E T} \right|^2.
\end{equation}
The kernel $K(T,E)$ is a partial Fourier transform of the connected two-point correlator, $\disAv{\rho(E_1)\rho(E_2)} - \disAv{\rho(E_1)}\disAv{\rho(E_2)}$. We also assume that the TRP can be obtained as the exponential of a stochastic matrix $M$, 
\begin{equation}
    \trp(T) = \tr(e^{MT}) = 1 + \sum_\ell e^{-\lambda_\ell T},
\end{equation}
where $M$ has one zero eigenvalue and some number of additional negative eigenvalues denoted $-\lambda_\ell$.

In terms of this data, the kernel $K$ is 
\begin{equation}
\begin{split}
    K(T,E)= K_{\textrm{short time}}(T,E)+ K_{\textrm{long time}}(T,E)\\
     K_{\textrm{short time}}(T,E)=\frac{|T|}{2\pi}\left(1+\sum_\ell e^{-\lambda_\ell|T|}\right)\\
     K_{\textrm{long time}}(T,E)=\frac{\lambda_1e^{-\lambda_1 |T|}}2*\frac{\lambda_2e^{-\lambda_2 |T|}}2*\frac{\lambda_3e^{-\lambda_3 |T|}}2...* K_{\textrm{long time}}^0(T,E)
\end{split}
\label{eq:fullSFF}
\end{equation}
The $*$s in equation \eqref{eq:fullSFF} refer to repeated convolution. $K_{\textrm{long time}}^0(T)$ is the RMT result for the kernel
\begin{equation}
    K_{\textrm{long time}}^0(T)=\begin{cases}
        0 & \text{if } |T|\leq 2\pi \bar\rho \\
        \bar \rho-\frac{|T|}{2\pi} & \text{if } |T|> 2\pi \bar\rho
    \end{cases}
\end{equation}
Note that the only potential sources of energy dependence in $K$ are the density of states $\bar{\rho}$ and the rates $\lambda_\ell$.

\begin{figure}
    \centering
    \includegraphics{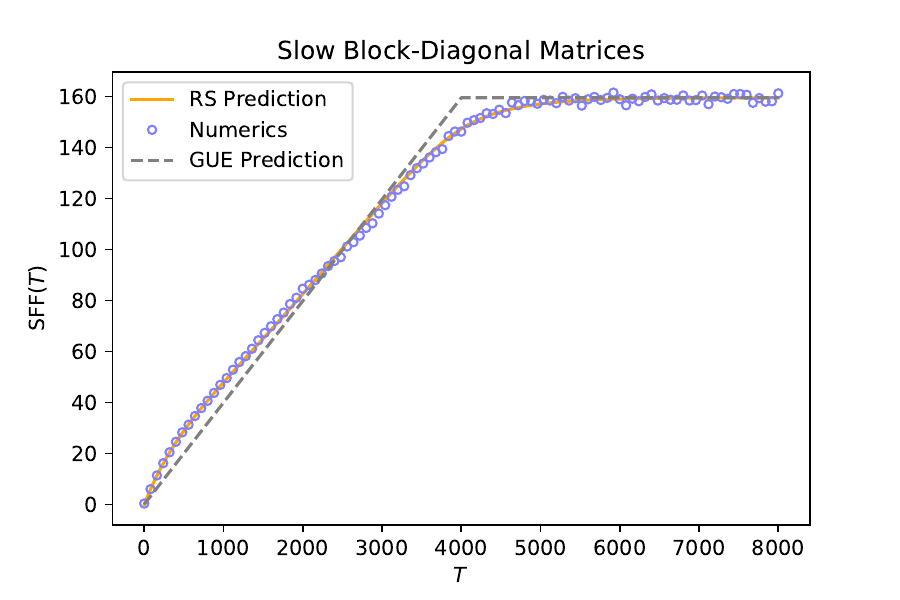}
    \caption{Our theory (orange line) is in excellent agreement with a direct numerical evaluation of the SFF for a GUE-like ensemble with two 1000-by-1000 blocks (blue circles). One can see clear deviations from the pure GUE result (grey dashed line) and a good agreement with \eqref{eq:fullSFF} at early and late time.}
    \label{fig:IntroPrediction}
\end{figure}

The fine print is that this equation applies when the Heisenberg time of the system is much longer than the Thouless time, $T_{\text{Th}} \sim (\min \lambda_\ell)^{-1}$. For a very different analysis that holds when the Thouless time is longer than the Heisenberg time (but which makes other assumptions) see \cite{RichardGlass}. Although we believe \eqref{eq:fullSFF} holds for all types of chaotic systems with the same symmetries as GUE matrices, the derivation below assumes a semiclassical limit. Beyond that we have numerical data confirming our hypothesis. One example of the behavior is shown in Figure \ref{fig:IntroPrediction}. One can see a clear if modest enhancement at early time followed by a corresponding suppression at late time relative to the GUE result. Our theory gives a satisfactory accounting of both the enhancement and the suppression. 

Rather remarkably, a special case of \eqref{eq:fullSFF} was derived almost 30 years ago in the context of electrons in a disordered metallic grain~\cite{1994JETPL..60..656K,1995PhRvL..75..902A} and again later using sigma model technology~\cite{2000PhRvL..85.5615A}. In our language, these works considered a situation where the TRP is controlled by a single particle diffusion process. Initially unaware of these works, we came to our result by a different route, trying to understand the late-time properties of the `block Rosenzweig-Porter' SFF considered in~\cite{RichardGlass}. In light of the earlier results, our contributions are as follows: (1) we conjecture that equation \eqref{eq:fullSFF} is general for systems with GUE-type symmetry, including for many-body systems, (2) we provide a derivation using periodic orbit theory and the lookalike, one which includes the diffusive case and more, (3) we emphasize the role of the sum rule and the general connection between early and late times that it implies, and (4) we generalize \eqref{eq:fullSFF} to the case of Floquet dynamics (Appendix~\ref{app:floquet}).

The rest of the paper is organized as follows. In Section \ref{sec:review}, we first review the spectral form factor and its relation to the total return probability. In Section \ref{sec:SFFInt}, we review and expound upon the sum rule first discussed in \cite{RichardGlass} and then show that \eqref{eq:fullSFF} obeys the sum rule. In Section \ref{sec:derivation}, we derive \eqref{eq:fullSFF} using semi-classical methods. In Section \ref{sec:outlook}, we conclude with an outlook. We set $\hbar =1$ throughout, so that the semi-classical limit is the limit of large dimensionless actions.

\section{Review of the Spectral Form Factor}
\label{sec:review}

We now review the basic properties of the spectral form factor (SFF) and the total return probability (TRP). As written above with $e^{-iHT}$s, the SFF at time $T$ looks like a diagnostic of dynamical behavior at time $T$. It can also be viewed as a Fourier transform of a two-point function of the density of states
\begin{equation}
    \SFF(T,f)=\disAv{\int dE_1dE_1 \rho(E_1)\rho(E_2) f(E_1)f(E_2)e^{i(E_1-E_2)T}},
\end{equation}
where, again, $\rho(E)=\sum_n \delta(E-E_n)$ is the density of states.

The $f$s are typically chosen as slowly-varying filter functions which restrict the spectral form factor to some small energy window of interest. A common choice would be a Gaussian $f(E)=\exp\left(-\frac{(E-E_0)^2}{4\sigma^2}\right)$ where $E_0$ is the energy scale we want to study and $1/\sigma$ is shorter than the time scales we are curious about but $\sigma$ is not enough energy to appreciably change the dynamics of the system, e.g. the energy diffusivity.

It is very useful to decompose the SFF into two parts, a connected SFF and a disconnected SFF. If we consider the random variable $\tr(f(H)e^{iHT})$, the disconnected SFF is the absolute square of the average,
\begin{equation}
    \SFF_{\text{disc}} = \left|\disAv{\tr(f(H)e^{iHT})}\right|^2.
\end{equation}
The connected SFF is the full SFF less the disconnected part,
\begin{equation}
    \SFF_{\text{conn}}(T,f)= \SFF - \SFF_{\text{disc}}={\int dE_1dE_1 \disAv{\rho(E_1)\rho(E_2)}_{\text{conn}} f(E_1)f(E_2)e^{i(E_1-E_2T)}},
    \label{eq:SFFcon}
\end{equation}
where $\disAv{\rho(E_1)\rho(E_2)}_{\text{conn}}=\disAv{\rho(E_1)\rho(E_2)}-\disAv{\rho(E_1)}\disAv{\rho(E_2)}$ is the connected two-point function of the density of states.

The connected SFF can be also expressed as
\begin{equation}
    \SFF_{\text{conn}}(T,f)=\int dE f^2(E) K(T,E),
\end{equation}
where
\begin{equation}
\begin{split}
    K(T,E) &=\int \frac{d\omega}{2\pi} e^{-i \omega T} K(\omega,E) \\
    K(\omega,E) &= 2 \pi \disAv{ \rho(E+\omega/2)\rho(E-\omega/2) }_{\text{conn}} 
\end{split}
\end{equation}
is a function whose short time behavior encodes details about the system's dynamics but whose long-time behavior seems to agree with random matrix theory for all ergodic systems. Going forward, we will usually suppress the $E$ and write $K(T,E)=K(T)$. For the GUE ensemble in random matrix theory, we have the famous result:
\begin{equation}
    K_0(T,E)=\begin{cases}
        \frac{|T|}{2\pi } & \text{if } |T|\leq 2\pi \bar\rho(E) \\
        \bar{\rho} & \text{if } |T|> 2\pi \bar\rho(E)
    \end{cases}
    \label{eq:KRMT}
\end{equation}

%\begin{figure}
%    \centering
%    \includegraphics[scale=0.5]{figures/EarlyRatio.pdf}
%    \caption{The SFF enhancement acts as a diagnostic of thermalization. The blue line is the numerical value of the SFF for a block random matrix system divided by the RMT prediction. The orange line is the total return probability, the sum over the blocks that a start started in that block will still be in that block after time $T$. We see that this measure of thermalization agrees perfectly with the SFF enhancement.}
%    \label{fig:my_label}
%\end{figure}

One particularly important link between spectral statistics and thermalization is the result of \cite{2020Prosen,Winer_2020,Roy_2022}, which extends equation \eqref{eq:KRMT} beyond pure random matrix theory. For our purposes, the Total Return Probability (TRP) is defined as follows: Partition the configuration space into various sectors indexed by $i,j$. Let $p_{i\to j}(T)$ be the probability that a system starting in sector $i$ at time $0$ has evolved to sector $j$ at time $T$. Then the TRP defined as 
\begin{equation}
\trp(T)=\sum_j p_{j\to j}(T).    
\end{equation}
The TRP is a natural measure of thermalization: if the system has a strong memory of the sector it started in, then the TRP is large, and if the system reaches equilibrium, then $p_{j\to j}(T)$ is just the equilibrium probability of $j$ and so the TRP is $1$.

Remarkably, the TRP is also reflected in the short time spectral statistics. For times on the scale of the equilibration time (assumed here to be much less than the Heisenberg time) we have
\begin{equation}
    K(T)= \trp(T) \frac{T}{2\pi}.
\end{equation}
%\begin{figure}
%    \centering
%    \includegraphics[scale=0.5]{figures/diagonal.pdf}
%    \caption{A pair of identicial paths contributing to the connected SFF.}
%    \label{fig:diagonal}
%\end{figure}
This formula can be derived in a number of different ways for a number of different systems, but most of these methods are built upon one spiritual pillar: the diagonal approximation. This method calculates the connected SFF by writing the evolution operators in equation \ref{eq:sffDef} as sums over closed paths. Since there are two evolution operators, the connected SFF will be a sum over pairs of paths. In the semiclassical limit, these paths terms will have large phase contributions coming from the difference in action between paths (in units with $\hbar = 1$). The diagonal approximation assumes that only pairs where right and left are the same contribute, while all other phases are random and on average add zero to the connected SFF. This approximation has had great success in calculating disorder-averaged SFFs, although there is additional work necessary in order to understand the behavior without disorder averaging \cite{halfWH,Garcia_Garcia_2022}.

As an example, if the system were a billiard with three weakly connected chambers, then those chambers define natural sectors. At early times, each chamber behaves likes its own independent system and the total Hamiltonian is like three independent random matrices with no inter-matrix level repulsion (SFF is $3\times$RMT), while at later times the chambers communicate and the Hamiltonian behaves like a single large random matrix (SFF is RMT). As another example, if the system had a local conserved charge, then each sector might be parameterized by the coarse-grained charge density at every point. At early times, the long-wavelength charge fluctuations are almost conserved and the system breaks into many sectors with weak inter-sector level repulsion. At later times, the all but the infinite wavelength charge fluctuations relax and the system behaves like one big random matrix.

\section{The Spectral From Factor Sum Rule}
\label{sec:SFFInt}

We now introduce the sum rule, first seen in \cite{RichardGlass}, which necessitates a late-time suppression in the SFF to compensate for the early time enhancement just discussed in Section~\ref{sec:review}.

Consider a quantum system with non-degenerate energy levels $E_n$ with $n=1,\cdots,\mathcal{D}$. At late times, the SFF with $f=1$ should approach the dimension of the Hilbert space, $\mathcal{D}$. This statement refers to the disorder averaged quantity since the non-averaged has an erratic time-dependence which continues to fluctuate wildly even at late time. More generally, the filtered SFF approaches the quantity 
\begin{equation}
    \mathcal{D}_f = \sum_n f(E_n)^2,
\end{equation}
which reduces to the Hilbert space dimension when $f=1$. It is therefore interesting to consider the value of 
\begin{equation}
    I(\infty)=\int_{-\infty}^\infty \{\SFF(T)-\mathcal{D}_f\}dT,
\end{equation}
or more generally, the integral
\begin{equation}
    I(T_0) = \int_{-\infty}^\infty \{\SFF(T)-\mathcal{D}_f\} \exp\left(- \frac{T^2}{2 T_0^2}\right) dT.
\end{equation}

This integral can be evaluated directly from the definition to give
\begin{equation}
    I(T_0) = \sqrt{2 \pi T_0^2} \sum_{n\neq m} f(E_n) f(E_m) \mathbb{E}\left[ \exp\left(- \frac{T_0^2 (E_n-E_m)^2}{2} \right) \right].
\end{equation}
It is also instructive to recast it in terms of the 2-point correlation as 
\begin{equation}
     I(T_0) = \sqrt{2 \pi T_0^2} \int dE d\omega f(E+\omega/2) f(E-\omega/2) \mathbb{E}\left[ \rho(E+\omega/2) \rho(E-\omega/2) \right]  \exp\left(- \frac{T_0^2 \omega^2}{2} \right) .
\end{equation}
Now, when $T_0$ is much larger than the Heisenberg time, the Gaussian in the integrand of $I$ functions like a delta function. In this limit, the integral becomes \begin{equation}
    \lim_{T_0\rightarrow \infty} I(T_0) = 2 \pi \int dE   f(E)^2 \lim_{\omega \rightarrow 0^+} \mathbb{E}\left[ \rho(E+\omega/2) \rho(E-\omega/2) \right].
\end{equation}
Remarkably, this limit is zero so long as there is any sort of level repulsion on any scale. This includes all the classic random matrix ensembles, as well Rosenzweig-Porter ensembles~\cite{Rosenzweig:1960zz} with arbitrarily weak off-diagonal coupling.

For chaotic systems with the dip-ramp-plateau structure, $\SFF - \mathcal{D}_f$ takes characteristic values during each phase of the dynamics. During the dip, it has an $O(\mathcal{D}^2)$ value for an $O(1)$ period of time. During the ramp, it has a negative $O(\mathcal{D})$ value for an $O(\mathcal{D})$ time. And as it enters the the plateau regime, it must go to zero. Thanks to the sum rule, we now know that the contributions from these regions precisely cancel out. This should hold for GOE, GUE, and GSE matrices. Furthermore, the shape of the dip more generally depends only on the density of states (and the energy window via $f$). One can show that total area under the disconnected SFF can be written as $2 \pi \int dE \bar{\rho}(E)^2 f(E)^2$. This quantity doesn't depend on the thermalization properties of the system, even implicitly (indeed, if the window is small enough it doesn't depend on any properties of the system except the disorder-averaged density of states in that window). Thus, if slow thermalization enhances the ramp at early times, so that $\SFF - \mathcal{D}_f$ is larger than the RMT result, then the integrand must be suppressed at some later time to `pay' back the enhancement. Since we just noted that the disconnected SFF is independent of the thermalization properties, it must be that this suppression appears in the connected SFF. As it turns out, for GUE-like models, the suppression appears to be localized around the Heisenberg time.

%%%

As a further exercise, let us check that our conjectured formula obeys the sum rule. Recall again that $\bar{\rho}(E) = \langle \rho(E)\rangle$ is the average density of states of our Hamiltonian. The Fourier transform of $K(T,E)$ is defined as
\begin{equation}
    K(\omega,E) = \int dT e^{i \omega T} K(T,E),
\end{equation}
and the inverse transform is 
\begin{equation}
    K(T,E) = \int \frac{d\omega}{2\pi} e^{- i \omega t} K(\omega,E).
\end{equation}
The necessary Fourier transforms are recorded and discussed in Appendix \ref{app:fourier}. 

Let us start with the connected 2-point function in the GUE ensemble,
\begin{equation}
    K_{\text{GUE}}(T)=\frac{1}{2\pi} \left( 2\pi \bar{\rho} + |T| -\frac{1}{2}|T+2\pi \bar{\rho}| -\frac{1}{2} |T-2\pi \bar{\rho}|\right).
\end{equation}
Note that we are suppressing the dependence on the ``center of mass'' energy $E$. In the frequency/energy domain, it can be written as
\begin{equation}
    K_{\text{GUE}}(\omega)=2\pi \bar{\rho}\delta(\omega) - \frac{1 - \cos(\omega 2\pi \bar{\rho})}{\pi \omega^2}.
\end{equation}
The sum rule is satisfied if and only if
\begin{equation}
    \lim_{\omega \rightarrow 0} \left\{ K(\omega) - 2 \pi \bar{\rho} \delta(\omega)  \right\} + 2\pi \bar{\rho}^2=0.
\end{equation} 

Let us first consider just the short-time correction,
\begin{equation}
    \Delta K_{\text{short time}} = \sum_\ell \frac{|T|}{2\pi} e^{-\lambda_\ell |T|} \subset K(T,E).
\end{equation}
Its Fourier transform is
\begin{equation}
    \Delta K_{\text{short time}}(\omega) = -\frac{1}{2 \pi} \sum_\ell \left[\frac{1}{(\omega+i \lambda_\ell)^2} + \frac{1}{(\omega-i \lambda_\ell)^2}\right],
\end{equation}
and its zero frequency limit is
\begin{equation}
    \Delta K_{\text{short time}}(\omega\rightarrow 0) = \frac{1}{\pi} \sum_\ell \frac{1}{\lambda_\ell^2} \neq 0.
\end{equation}
Hence, since the pure RMT $K$ obeys the sum rule, including only this early time correction into the SFF violates the sum rule.

Now let us investigate the effect of the late time correction. Using the notation of \eqref{eq:fullSFF}, the Fourier transform of $K_{\text{long time}}$ is
\begin{equation}
    \left(2\pi \bar{\rho} \delta(\omega) + \frac{\cos(2\pi \omega \bar{\rho})}{\pi \omega^2} \right) \prod_\ell \frac{\lambda_\ell^2}{\omega^2 + \lambda_\ell^2} = 2\pi \bar{\rho} \delta(\omega) + \frac{\cos(2\pi \omega \bar{\rho})}{\pi \omega^2} \prod_\ell \frac{\lambda_\ell^2}{\omega^2 + \lambda_\ell^2} ,
\end{equation}
where we understand the double pole at $\omega=0$ as an equal average of a double pole just below the real axis and a double pole just above the real axis. Note that the $2\pi \bar{\rho} \delta(\omega)$ term is unmodified by the inclusion of the product over $\lambda_\ell$s. The correction at long time is thus
\begin{equation}
    \Delta K_{\text{long time}}(\omega) = \frac{\cos(2\pi \omega \bar{\rho})}{\pi \omega^2}\left[\left( \prod_\ell \frac{\lambda_\ell^2}{\omega^2 + \lambda_\ell^2} \right) -1 \right]
\end{equation}
The zero frequency limit of the correction is
\begin{equation}
    \Delta K_{\text{long time}}(\omega \rightarrow 0) = -\frac{1}{\pi} \sum_\ell \frac{1}{\lambda_\ell^2} + O(\omega^2).
\end{equation}
Remarkably, this perfectly cancels the enhancement from the short-time correction to ensure the sum rule is satisfied.

Our goal in the remainder of this paper is thus to establish the formula \eqref{eq:fullSFF}, or its Fourier equivalent,
\begin{equation}
    K(\omega) = 2\pi \bar{\rho}\delta(\omega) - \frac{1}{\pi \omega^2 }-\frac{1}{2 \pi} \sum_\ell \left[\frac{1}{(\omega+i \lambda_\ell)^2} + \frac{1}{(\omega-i \lambda_\ell)^2}\right] + \frac{\cos(2\pi \omega \bar{\rho})}{\pi \omega^2}  \prod_\ell \frac{\lambda_\ell^2}{\omega^2 + \lambda_\ell^2}.
    \label{eq:KRMT_F}
\end{equation}

%%%
\section{A Derivation from Periodic Orbit Theory}
%%%
\label{sec:derivation}

We have shown that our conjectured formula obeys the sum rule, but this constraint alone is not sufficient to single out the formula \eqref{eq:fullSFF}. To give a complete argument, we need a toolbox that gives access to both short and long times and which can accommodate the effects of a non-trivial total return probability (TRP). The periodic orbit theory of semi-classical quantum chaos turns out to be well suited to the task.

In this section, we thus use periodic orbit theory to derive \eqref{eq:fullSFF} and \eqref{eq:KRMT_F}. We emphasize that we have ample numerical evidence that \eqref{eq:fullSFF} and \eqref{eq:KRMT_F} work well beyond systems with a semi-classical limit, but this limit does provide an extremely useful set of tools. In Section~\ref{sec:outlook}, we discuss what would be needed to give a more general proof that does not rely on semi-classical methods.

This section is organized as follows. First, we review the spectral determinant, the Gutzwiller trace formula, and the Berry-Keating Riemann-Siegel lookalike. Second, since the lookalike formula is key, we review the Riemann-Siegel formula for the Riemann zeta function upon which the lookalike is based. Third, we use the lookalike formula to directly obtain the Fourier transform of \eqref{eq:fullSFF} given in \eqref{eq:KRMT_F}.

%%%
\subsection{The Spectral Determinant}
%%%

In this section we will review some properties of a function of energy called the spectral determinant whose key property is that it has zeros at the energy eigenvalues of the system. We will also quickly review the necessary aspects of the periodic orbit approach to quantum chaos which gives a formula for te spectral determinant. Nothing in this section is new, but it will be important setup for the derivation of \eqref{eq:KRMT_F}. 

We start with a Hamiltonian $H$ with energy levels $E_n$ and define the spectral determinant
\begin{equation}
    \Delta(E)=\prod_n A(E,E_n) (E-E_n),
    \label{eq:specDet}
\end{equation}
where $A$ is some slowly varying function with no zeroes, chosen to make the product in \eqref{eq:specDet} converge. For instance, for a 1D particle in a box $A=\frac 1E$ would work, while for a 1D particle in a harmonic trap that product would diverge but $A=\frac{E+E_n}{E^2}$ would converge.

The determinant \eqref{eq:specDet} is closely related to the resolvent,
\begin{equation}
    R(E)=\tr \frac{1}{E-H}\approx \frac{d \log \Delta (E)}{d E} \label{eq:resolveDet},
\end{equation}
where we use $\approx$ instead  of $=$ because of the slowly varying $A$ function.

If $H$ has a semiclassical limit, we can derive an expression for $\Delta$ using the Gutzwiller trace formula. Let us first recall the idea of the trace formula for the resolvent. Including a small imaginary part, the resolvent can be written
\begin{equation}
    \tr\frac{1}{E + i \epsilon - H  } = \int_0^\infty dt e^{i (E + i\epsilon ) t} \tr e^{- i H t}.
\end{equation}
Owing to the trace, the quantity $\tr e^{- i H t}$, when expressed as a path integral, will be controlled by periodic orbits of the corresponding classical equations of motion. The integral over $t$ then yields a sum over periodic orbits of various periods. One must also include the functional determinant describing the fluctuations around each classical orbit that contributes to the path integral. This determinant turns out to depend on the stability properties of the orbit, which should not be surprising since it concerns small fluctuations around the periodic orbit.

The result of this line of thinking is the famous Gutzwiller trace formula,
\begin{equation}
    \tr \frac{1}{E + i \epsilon -H} \sim - i \pi \bar{\rho}(E) - i \sum_a T_a F_a e^{iS_a},
\end{equation} 
where $a$ indexes the periodic orbits, $T_a$ is the period of the orbit, $F_a$ is a dimensionless quantity related to the stability of the orbit, and $S_a$ is the classical action of the orbit. Note that one has two different ways of regulating the poles by moving them off the real axis. The quantities $F_a$ are essentially return amplitudes, a fact which we will use to relate them to the TRP below. We will also assume that they are slowing varying functions of energy. We use the $\sim$ in the statement of the trace formula since the sum is not obviously converent and the resolvent itself typically requires regularization in a system with unbounded spectrum.

Using the relationship between the resolvent and the spectral determinant, \eqref{eq:resolveDet}, one gets a corresponding formula for the spectral determinant,
\begin{equation}
    \Delta (E + i \epsilon )=B(E) \exp \left(-i \pi \bar{\mathcal  N} (E)-\sum_a F_a e^{iS_a}\right),
    \label{eq:DeltaProd}
\end{equation}
where $B(E)$ is a smooth, slowly varying function reflecting the impact of the $A$ regulator and $\bar {\mathcal N}(E)$ is the smoothed out number of states below energy $E$. It can be defined using the Heaviside step function as
\begin{equation}
    \bar{\mathcal  N} (E) = \int \frac{d^dx d^dp}{\hbar^d} \theta(E-H_{\text{cl}}(x,p)),
\end{equation}
where $2d$ is the dimension of phase space. The smoothed density also the property that $\frac{d\bar{\mathcal  N}(E)}{dE}=\bar \rho(E)$. The first term in the exponent in \eqref{eq:DeltaProd} is a straightforward integration of $- i \pi \bar{\rho}$. The second term in the exponent is the integral of the erratic second term assuming $F_a$ depends weakly on energy and using $\frac{\partial S_a}{\partial E} = T_a$.

Remember also that there were two ways of regulating the resolvent with an $i \epsilon$ prescription. The other choice leads to a naively different formula for the spectral determinant,
\begin{equation}
    \Delta (E - i \epsilon )=B(E) \exp \left(i \pi \bar{\mathcal  N} (E)-\sum_a F_a^* e^{-iS_a}\right).
    \label{eq:DeltaProd_alt}
\end{equation}
However, the spectral determinant is non-singular (it has zeros at the physical energies, not poles), so the two different $i \epsilon$ procedures should agree. This is certainly not obvious from the from of $\Delta(E \pm i \epsilon)$, and indeed one motivation for the lookalike formula discussed below is to make this equality manifest.

Equation \eqref{eq:DeltaProd} has a sum in an exponential, and can thus be interpreted as an infinite product with an overall phase factor. It can be rewritten as a sum,
\begin{equation}
    \Delta (E)=B(E) e^{-i \pi \bar{\mathcal  N} (E)}\sum_A (-1)^{|A|}F_A e^{iS_A},
    \label{eq:DeltaSum}
\end{equation}
where each $A$ represents a `pseudo-orbit' which is a combination of $|A|$ periodic orbits $a \in A$. The action is $S_A = \sum_{a \in A} S_a$ where the sum may have multiplicity. Similarly, $F_A$ is a product over $F_a$s (with combinatorial factors of $1/k!$ when the same semiclassical orbit is included $k$ times). 

We can also invert equation \eqref{eq:DeltaProd} or \eqref{eq:DeltaProd_alt} to get
\begin{equation}
\begin{split}
        \Delta (E + i\epsilon)^{-1}=B(E)^{-1} e^{+ i \pi \bar{\mathcal  N} (E)}\sum_A F_A e^{+ iS_A},\\
        \Delta (E - i\epsilon)^{-1}=B(E)^{-1} e^{- i \pi \bar{\mathcal  N} (E)}\sum_A F_A^* e^{- iS_A}.
        \label{eq:deltaInv}
\end{split}
\end{equation}
It is interesting that even though $\Delta$ and $\Delta^{-1}$ have extremely different properties (for instance $\Delta$ is analytic on the real line whereas $\Delta^{-1}$ has a series of poles) the formulas in equations \eqref{eq:DeltaSum} and \eqref{eq:deltaInv} look so similar.

Now, all these expressions are quite formal at present, for example, the reality of the resolvent and the spectral determinant are not manifest. The convergence of the potentially infinite sum over periodic orbits is also far from clear. Indeed, it turns out that as the orbits and pseudo-orbits get longer, their number proliferates exponentially as $e^{\lambda_L T}$, where $\lambda_L$ is a Lyapunov exponent. The amplitudes $F_a$ decrease merely as $e^{-\lambda_L T/2}$, meaning that the sums aren't even absolutely convergent.

This confusing situation was clarified by Berry and Keating in a remarkable series of papers~\cite{10.1007/3-540-17171-1_1,10.2307/52022,10.2307/52059,MVBerry_1990}. Inspired by the Riemann-Siegel resummation formula for the Riemann zeta function, they proposed a resummation of the periodic orbit theory which addresses the problems outlined above. Their formula is
 \begin{equation}
    \Delta (E)=B(E) e^{-i \pi \bar{\mathcal  N} (E)}\sum_{T_A<T_H/2} (-1)^{|A|}F_A e^{iS_A}+B(E) e^{i \pi \bar{\mathcal  N} (E)}\sum_{T_A<T_H/2} (-1)^{|A|}F_A^* e^{-iS_A},
    \label{eq:Lookalike}
 \end{equation}
 where now one restricts to pseudo orbits $A$ of total length less than half the Heisenberg time, $T_H = 2 \pi \hbar \bar{\rho}$. Among other virtues, this formula for $\Delta(E)$ is now manfestly real since the two terms are complex conjugates. This formula is called the Riemann-Siegel lookalike. 

%%%
\subsection{The Riemann-Siegel Formula and its Lookalike}
%%%

In order to understand the resummation Berry and Keating did for $\Delta(E)$, it is helpful to start with a simpler case: the resummation done over a hundred years ago for the Riemann zeta function.

We will begin with a review of the properties of the Riemann function $\zeta(s)$ \cite{edwards1974riemann,montgomery2017exploring}. It is most commonly defined by a so-called Dirichlet series,
\begin{equation}
    \zeta(s)=\sum_{n=1}^\infty \frac 1{n^s},
    \label{eq:zetaSum}
\end{equation}
for $\textrm{Re } s>1$ and by analytical continuation for the rest of the complex plane. Even before Riemann, it was shown by Euler\cite{apostol1998introduction} that $\zeta(s)$ could be rewritten 
\begin{equation}
    \zeta(s)=\prod_{\textrm{prime numbers }p} \frac {p^s}{p^s-1}. \label{eq:zetaProd}
\end{equation}
This identity was the basis for the discovery of several new facts about prime numbers. For instance, the fact that equation \eqref{eq:zetaProd} diverges when $s=1$ (the only pole of $\zeta(s)$) was the first new proof of the infinitude of the primes in over two thousand years, and lent strength to the first estimates of the density of prime numbers. 

Riemann used Poisson resummation to prove that 
\begin{equation}
    \xi(s)\equiv\pi^{-\frac{s}{2}}\Gamma\left(\frac s2\right)\zeta(s)=\pi^{-\frac{1-s}{2}}\Gamma\left(\frac {1-s}2\right)\zeta(1-s)\equiv\xi(1-s).
\end{equation}
This is the famous reflection formula, called that because the same function is applied to both the left and right but reflected about the point $s=\frac 12$. We can plug in $s=\frac 12+it$ to get
\begin{equation}
    \xi\left(\frac 12+it\right)=\xi\left(\frac 12-it\right)=\xi\left(\frac 12+it\right)^*,
\end{equation}
where the first equality is the reflection formula and the second equality follows from the definition. Thus $\xi$ is real on the line $s=\frac 12+it$. This so-called `critical line' has several other important properties, the most famous being that it is the conjectured location of all of the $\zeta$ function's nontrivial zeroes. This conjecture, called the Riemann hypothesis, is one of the most famous open questions in mathematics, with implications in analytic number theory and beyond.

Due to the importance of the Riemann hypothesis, there was has been substantial interest in numerically calculating the zeta function high on the critical line. Using either the sum in equation \eqref{eq:zetaSum} or the product in equation \eqref{eq:zetaProd} would give an absolutely divergent but highly oscillating result. Such sums are technically difficult to deal with.

Unbeknownst to the wider mathematical world, the solution to these difficulties was lying in Bernhard Riemann's notes for over 70 years. It was discovered in those notes by mathematical Carl Siegel, who added his own technical contributions. The result was the Riemann-Siegel formula~\cite{zbMATH02552973}.

This formula is best stated as calculating not $\zeta$ but the related function
\begin{equation}
\begin{split}
    Z(t)=\zeta\left(\frac 12+it\right) e^{i\theta(t)}\\
    \theta(t)=\textrm{arg}\left[\Gamma\left(\frac 14 +\frac{it}{2}\right)\right]-\frac{\log \pi}{2}t \approx \frac t2 \log \frac{t}{2\pi}-\frac t2-\frac \pi 8;
\end{split}
\end{equation}
$\theta(t)$ is the Riemann-Siegel theta function. Using the reality of $\xi$ on the critical line, $Z(t)$ is easily shown to be real. But this reality is hardly obvious when looking at the formula. The situation is analogous to the case with above with the periodic orbit expansion, with $\theta(t)$ playing the role of $\pi \bar{\mathcal{N}}$ and $Z(t)$ playing the role of the spectral determinant.

Using a resummation technique, it is posssible to produce a series approximation to $Z(t)$ that improves the convergence properties and makes the reality manifest. The starting point is to write
\begin{equation}
    \begin{split}
    Z(t) & =e^{i\theta(t)}\sum_{n=1}^\infty \frac{1}{n^{\frac 12+it}}\\
    & = e^{i\theta(t)}\sum_{n=1}^N \frac{1}{n^{\frac 12+it}}+e^{i\theta(t)}\sum_{n=N+1}^\infty \frac{1}{n^{\frac 12+it}}\\
    & = e^{i\theta(t)}\sum_{n=1}^N \frac{1}{n^{\frac 12+it}}+e^{i\theta(t)}\sum_{k=-\infty}^\infty\int_{n=N+\epsilon}^\infty e^{2\pi i k n}\frac{1}{n^{\frac 12+it}}dn.
    \end{split}
\end{equation}
So far $N$ is undetermined and everything is exact. The approximation enters by performing the integrals on the final line via the method of stationary phase. For the term labelled by $k$, the stationary point is $n_*=\frac{t}{2\pi k}$. By choosing  $N=\sqrt{\frac{t}{2\pi}}$, we find that the first $N$ terms of the second sum have their saddle points in the interval $n \in [N,\infty]$. In fact, the resulting terms are exactly the conjugates of the terms in the first sum.\footnote{This equality requires including the quadratic fluctuations around the saddle point.} Thus we have 
\begin{equation}
    Z(t)\approx e^{i\theta(t)}\sum_{n=1}^N \frac{1}{n^{\frac 12+it}}+e^{-i\theta(t)}\sum_{n=1}^N \frac{1}{n^{\frac 12-it}} \label{eq:RiemannSiegel}
\end{equation}
For $t$s where $\sqrt{\frac t{2\pi}}$ takes on integer values, equation \eqref{eq:RiemannSiegel} is very close to exact, while for intermediate $t$s some interpolation is necessary.

Continuing with the analogy to the periodic orbit theory, $Z(t)$ can be thought of as a spectral determinant analogous to $\Delta(E)$. Both functions are real, both can be expressed as an overall phase ($\theta$ or $\bar {\mathcal N}$) times either a sum (equations \eqref{eq:zetaSum} and \eqref{eq:DeltaSum}) or a product (equations \eqref{eq:zetaProd} and \eqref{eq:DeltaProd}). Furthermore, Montgomery's pair correlation conjecture \cite{montgomery1973,Goldston1987} states that the zeroes of $Z$ repel each other just like the zeroes of $\Delta$ in a chaotic system with no time-reversal. A great deal of effort \cite{leboeuf2001riemannium,riemannReview} has been put into trying to find a specific Hamiltonian system whose spectral determinant corresponds to $Z$, thus far with no success.

 A more successful program, however, has been the adaptation of equation \eqref{eq:RiemannSiegel} to Hamiltonian systems, which we gave above as \eqref{eq:Lookalike} and which we repeat here:
 This adaptation is 
 \begin{equation}
    \Delta (E)=B(E) e^{-i \pi \bar{\mathcal  N} (E)}\sum_{T_A<T_H/2} (-1)^{|A|}F_A e^{iS_A}+B(E) e^{i \pi \bar{\mathcal  N} (E)}\sum_{T_A<T_H/2} (-1)^{|A|}F_A^* e^{-iS_A}.
 \end{equation}
 This Riemann-Siegel `lookalike' has been shown to hold in systems as diverse as Floquet maps~\cite{Braun_2012} and quantum graphs~\cite{Waltner_2019}. In complete analogy to the above manipulations with the Riemann zeta function, Berry and Keating argued for the lookalike based on the relationships between long and short periodic orbits and some generalization of Poisson resummation. Additionally, as we review below, this lookalike formula also makes it straightforward to obtain both the ramp and the plateau via a simple `diagonal approximation'. The reader will recall that such an ability to address both the ramp and the plateau was one of our central desiderata. We will therefore assume the lookalike and show that our main result follows.

%%%
\subsection{The Spectral Form Factor from the Riemann-Siegel Lookalike}
%%%

\label{sec:SFFresult}
In this section we will extract an expression for the spectral form factor from spectral determinants. We want to use spectral determinants because this gives access to the lookalike. We will largely follow the logic in \cite{Keating_2007}.

The spectral form factor is a two-point function of the density of states, and the density of states is the logarithmic derivative of $\Delta$. Thus we define
\begin{equation}
    Q(\alpha,\beta,\gamma,\delta)=\frac{B(E+\alpha)B(E-\beta)}{B(E+\gamma)B(E-\delta)}\disAv{ \frac{\Delta(E+\gamma)\Delta(E-\delta)}{\Delta(E+\alpha)\Delta(E-\beta)} }.
    \label{eq:Qdef}
\end{equation}
Importantly, we will give $\alpha$ and $\beta$ small positive imaginary parts, $\alpha,\beta \rightarrow \alpha,\beta  + i \epsilon$. Following \cite{Haake} (Section 10.6), we can express two-point function of density in terms of $Q$:
\begin{equation}
    \disAv{\rho(E+\omega/2)\rho(E-\omega/2) }_{\text{conn}}=-\frac{1}{2\pi^2}\left[\textrm{Re} \frac{\partial^2}{\partial \alpha \partial \beta}Q \right]_{\alpha,\beta , \gamma,\delta = \omega/2}-\frac 12 \bar \rho^2.
    \label{eq:rho2Point}
\end{equation} 
This formula is discussed in Appendix~\ref{app:Qderiv}, and we review below how the usual RMT result is recovered from it by invoking the diagonal approximation and the lookalike. Then we show how to incorporate slow modes into the framework to obtain out main result. Finally, note that $Q$ in \eqref{eq:Qdef} is left invariant by the symmetry transformation
\begin{equation}
    \{\gamma \to -\delta, \delta \to -\gamma\}.
    \label{eq:switchSym}
\end{equation} 
There is no analogous symmetry for $\alpha,\beta$ because both have infinitesimal positive imaginary parts.

Given a way to evaluate equation \eqref{eq:Qdef}, and thus equation \eqref{eq:rho2Point}, a simple Fourier transform gives the SFF. Our strategy, then, will be to plug equations \eqref{eq:Lookalike} and \eqref{eq:deltaInv} into equation \eqref{eq:Qdef} in order to derive an expression for the SFF. The numerators are dealt with using \eqref{eq:Lookalike}, for a total of $2 \times 2$ groups of terms, while the denominators are dealt with using \eqref{eq:deltaInv}. Each group contains a four-fold sum over quasi-orbits. Two of these four groups of terms will contain rapidly oscillating phase factors, $e^{ \pm 2 i S}$, and will not contribute appreciably after the disorder averaging. The dropping these terms is part of the `diagonal approximation', and we will drop additional terms below by the same logic. 

\begin{figure}
    \centering
    \includegraphics[scale=1.2]{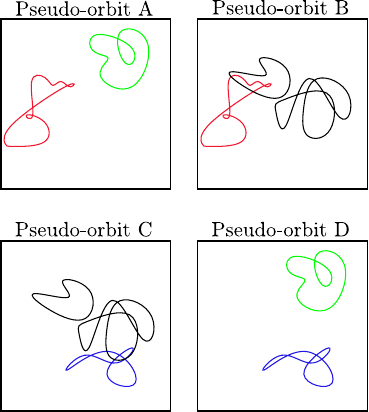}
    \caption{Four pseudo-orbits satisfying $A\cup C=B\cup D$. $A\cap B$, $C\cap D$ and $A\cap D$ each consist of one orbit, while $B \cap C$ consists of the two black orbits.}
    \label{fig:pseudo}
\end{figure}

The denominator contributes
\begin{equation}
 \sum_{AB} F_A F_B^* e^{i \pi \bar{\mathcal{N}}(E+\alpha) - i \pi \bar{ \mathcal{N}}(E-\beta)} e^{i S_A(E+\alpha) - i S_B(E-\beta)},   
\end{equation}
and the numerator contributes 
\begin{equation}
    \sum_{CD, T < T_H/2} F_C F_D^* (-1)^{|C|+|D|} e^{-i \pi \bar{\mathcal{N}}(E+\gamma) + i \pi \bar{ \mathcal{N}}(E-\delta)}e^{i S_C(E+\gamma) - i S_D(E-\delta)}   + \{\gamma \to -\delta, \delta \to -\gamma\}.
\end{equation}
And again, we have dropped terms for which the phases associated with the actions cannot cancel out, e.g. terms proportional $e^{i S_A - i S_B + i S_C + i S_D}$. At this point, we must ask which choices of $ABCD$ can give a significant contribution to the sum. The rapidly oscillating terms are
\begin{equation}
    e^{i \pi \left[ \bar{\mathcal{N}}(E+\alpha) -  \bar{ \mathcal{N}}(E-\beta) - \bar{\mathcal{N}}(E+\gamma) +  \bar{ \mathcal{N}}(E-\delta) \right]} e^{i \left[ S_A(E+\alpha) - S_B(E-\beta) +  S_C(E+\gamma) -  S_D(E-\delta) \right]} 
\end{equation}
and the corresponding $\{\gamma \to -\delta, \delta \to -\gamma\}$ term. Based on this expression, the rapidly oscillating terms can only cancel when the quasi-orbits satisfy
\begin{equation}
    A \cup C = B \cup D \,\,\,\, (\text{displayed term})
\end{equation}
or 
\begin{equation}
    A \cup D = B \cup C \,\,\,\, (\{\gamma \to -\delta, \delta \to -\gamma\} \text{ term}).
\end{equation}
This condition is illustrated in Figure~\ref{fig:pseudo}.

The next step is to expand $\bar{\mathcal{N}}(E+x)$ and $S_A(E+x)$ for generic small $x$, giving
\begin{equation}
    \bar{\mathcal{N}}(E+x) = \bar{\mathcal{N}}(E) + x \bar{\rho}(E) + \cdots
\end{equation}
and
\begin{equation}
    S_A(E+x) = S_A(E) + x T_A(E) + \cdots
\end{equation}
where, again, $\bar{\rho}(E)$ is the smooth or average density of states and $T_A(E)$ is the sum of the periods of the orbits in $A$. The resulting form of $Q$ is
\begin{equation}
\begin{split}
    Q= & e^{ i \pi \bar{\rho}(E) \{ \alpha + \beta - \gamma - \delta\}} 
 \sum_{ABCD} \mathbb{E}\bigg[ F_A F_B^* F_C F_D^* e^{i T_A \alpha + i T_B \beta + i T_C \gamma + i T_D \delta } \bigg] \\
    &+ \{\gamma \to -\delta, \delta \to -\gamma\}.
\end{split}
\label{eq:QuadProduct}
\end{equation}

Now, because the number of orbits proliferates so rapidly, it is a very good approximation to say that each orbit $a$ appears in a pseudo-orbit $A$ at most one time. In other words, we can safely neglect multiple appearances of a single periodic orbit. With this further approximation, which simplifies the form of the $F_A$s by removing certain combinatorial factors, it is now possible to rewrite \eqref{eq:QuadProduct} in an even simpler form. Consider the displayed term in \eqref{eq:QuadProduct} with $A \cup C = B \cup D$. We can break $A$ up into $A \cap B$ and $A \cap D$ and similarly $C$ into $C \cap B$ and $C \cap D$. Very conveniently, all the pieces of \eqref{eq:QuadProduct} interact well with this decomposition,
\begin{equation}
   F_A F_B^* F_C F_D^* = F_{A \cap B} F_{A \cap D} F_{B \cap A}^* F_{B \cap C}^* F_{C \cap B} F_{C \cap D } F_{D \cap A}^* F_{D \cap C}^* = |F_{A \cap B}|^2 |F_{A \cap D}|^2 |F_{C \cap B}|^2 F_{C \cap D}|^2,
\end{equation}
\begin{equation}
    (-1)^{|C|+|D|} = (-1)^{|B \cap C| + |C \cap D| + | A\cap D | + |C \cap D|} = (-1)^{|B \cap C| + |A \cap D|},
\end{equation}
and
\begin{equation}
    T_A \alpha + T_B \beta + T_C \gamma + T_D \delta = T_{A \cap B} (\alpha + \beta) + T_{A \cap D}(\alpha + \delta) + T_{C \cap B}(\gamma + \beta) + T_{C \cap D}(\gamma + \delta).
\end{equation}

The expression for $Q$ now refactorizes into pieces associated with $A\cap B$, $A \cap D$, $ C \cap B$, and $C \cap D$,
\begin{equation}
    \begin{split}
        Q = &  e^{ i \pi \bar{\rho}(E) \{ \alpha + \beta - \gamma - \delta\}} \mathbb{E}\Bigg[\sum_{A \cap B} |F_{A\cap B}|^2 e^{i T_{A\cap B} (\alpha + \beta)} \sum_{A \cap D} |F_{A\cap D}|^2 (-1)^{|A \cap D|} e^{i T_{A\cap D} (\alpha + \delta)} \\
        & \times \sum_{C \cap B} |F_{C \cap B}|^2 (-1)^{|C \cap B|} e^{i T_{C \cap B}(\beta + \gamma)} \sum_{C \cap D} |F_{C \cap D}|^2 e^{i T_{C \cap D} (\gamma + \delta)} \Bigg]\\
       & + \{\gamma \to -\delta, \delta \to -\gamma\}.
    \end{split}
\end{equation}

Remarkably, it can now be argued that the result is insenstive to removing the restriction that $T_C,T_D$ be less than $T_H/2$ (recall that this restriction was part of the lookalike formula). An example of this logic is given in \cite{Keating_2007} (Section 3). Finally, we define
\begin{equation}
\G(s)=\sum_A |F_A|^2(-1)^{n_A}{e}^{-sT_A}=\exp \left(-\sum_a |F_a^2|e^{-sT_a}\right),
\end{equation}
from which it follows that 
\begin{equation}
\G^{-1}(s)=\sum_A |F_A|^2{e}^{-sT_A}.
\end{equation}

One virtue of this series of manipulations is that the function $G(s)$ is approximately self-averaging, meaning we can at last carry out the ensemble average and write
\begin{equation}
\begin{split}
    Q(\alpha,\beta,\gamma,\delta)& =e^{\pi i \bar \rho (\alpha+\beta-\gamma-\delta)} \frac{\G(-i(\alpha+\delta))\G(-i(\beta+\gamma))}{\G(-i(\alpha+\beta))\G(-i(\gamma+\delta))}+\{\gamma \to -\delta, \delta \to -\gamma\}\\
    & =     e^{\pi i \bar \rho (\alpha+\beta-\gamma-\delta)}\frac{\G(-i(\alpha+\delta))\G(-i(\beta+\gamma))}{\G(-i(\alpha+\beta))\G(-i(\gamma+\delta))}+
    e^{\pi i \bar \rho (\alpha+\beta+\gamma+\delta)} \frac{\G(-i(\alpha-\gamma))\G(-i(\beta-\delta))}{\G(-i(\alpha+\beta))\G(i(\gamma+\delta))}.
\end{split}
    \label{eq:bigKeatingResult}
\end{equation}
Note that we obtain a formula which respects the symmetry in \eqref{eq:switchSym} using only the lookalike formula and the diagonal approximation. 

What does equation \eqref{eq:bigKeatingResult} tell us about $\disAv{ \rho(E+\omega/2)\rho(E-\omega/2) }_{\text{conn}}$? Returning to equation \eqref{eq:rho2Point}, which we repeat here,
\begin{equation}
    \disAv{\rho(E+\omega/2)\rho(E-\omega/2) }_{\text{conn}}=-\frac{1}{2\pi^2}\left[\textrm{Re} \frac{\partial^2}{\partial \alpha \partial \beta}Q \right]_{\alpha,\beta , \gamma,\delta = \omega/2}-\frac 12 \bar \rho^2,
\end{equation}
we see that two kinds of terms can arise in the derivative. One is a slowly oscillating term coming from the first of the terms in \eqref{eq:bigKeatingResult}. The other is a rapidly oscillating term going like $e^{2\pi i \omega \bar \rho}$ coming from the second of the terms in \eqref{eq:bigKeatingResult}. To say more, we need to understand the structure of $G(s)$.

We will discuss the properties of $\G$ in more detail below, but for now it suffices to state that $\G(s)$ has a simple root as $s$ goes to zero~\cite{Haake}. This means that when $\alpha=\beta=\gamma=\delta=\omega/2$, the numerator of the second term in \eqref{eq:bigKeatingResult} contains a double root. To get a nonzero answer, we need $\partial_\alpha$ in \eqref{eq:rho2Point} to act on $\G(-i(\alpha-\gamma))$ and $\partial_\beta$ in \eqref{eq:rho2Point} to act on $\G(-i(\beta-\delta))$. Taking $G(s)\propto s$ gives
\begin{equation}
  \disAv{\rho(E+\omega/2)\rho(E-\omega/2) }_{\text{conn}} \supset \frac {\cos 2\pi \omega \hat \rho}{2\pi^2\omega^2},
\end{equation}
which is the oscillating part that provides the famous plateau of the spectral form factor.

The full result for $Q$ assuming $G(s) \propto s$ is
\begin{equation}
    Q = e^{i \pi \bar{\rho}(\alpha+\beta-\gamma-\delta)} \frac{(\alpha+\delta)(\beta+\gamma)}{(\alpha+\beta)(\gamma+\delta)} -  e^{i \pi \bar{\rho}(\alpha+\beta+\gamma+\delta)}   \frac{(\alpha-\gamma)(\beta-\delta)}{(\alpha+\beta)(\gamma+\delta)}.
\end{equation}
As we said, the second term only gives a non-vanishing contribution when both derivatives act on the numerator. For the first term, the derivatives give
\begin{equation}
   \left[ \partial_\alpha \partial_\beta Q_{\text{first term}}\right]_{\alpha,\beta,\gamma,\delta \rightarrow \omega/2} = (i \pi \bar{\rho})^2 + \frac{1}{\omega^2},
\end{equation}
so both terms together give
\begin{equation}
   -\frac{1}{2\pi^2}\left[\textrm{Re} \frac{\partial^2}{\partial \alpha \partial \beta}Q \right]_{\alpha,\beta, \gamma,\delta = \omega/2} =  \frac{\bar{\rho}^2}{2} - \frac{1 - \cos 2\pi \omega \bar{\rho}}{2 \pi^2 \omega^2}.
\end{equation}

Now we turn at last to including slow modes into this description. To do so, we need to use the fact that the $F_a$s are return amplitudes, so the $|F_a|^2$s are return probabilities. Using the Hannay-Ozaria del Almeida theorem~\cite{Hannay_1984, haake2010quantum}, we can write
\begin{equation}
    \sum_{a:T<T_a<T+\Delta T}|F_a|^2 \approx \trp(T) \frac{\Delta T}{T},
\end{equation}
where $\trp$ is to the total return probability introduced earlier.

Recall that we assume a stochastic matrix ansatz for the TRP, 
\begin{equation}
    \trp = \tr(e^{MT}) = 1 + \sum_\ell e^{- \lambda_\ell T} .
\end{equation}
Plugging this into our expression for $G(s)$ yields
\begin{equation}
    \log G(s) = - \int^\infty_{T_{\text{short}}} dT \frac{\trp(T) }{T} e^{- s T} = - \int^\infty_{T_{\text{short}}} dT \frac{1 + \sum_\ell e^{-\lambda_\ell T}}{T} e^{- sT},
\end{equation}
where $T_{\text{short}}$ is a short-time cutoff below which our formula for the TRP fails. The integral has a logarithmic divergence as $s=0$ and whenever $s=-\lambda_\ell$, thus we may evaluate the integral as
\begin{equation}
    \log G(s) = - \log \frac{1}{s} - \sum_\ell \log \frac{1}{s+ \lambda_\ell} + \cdots,
\end{equation}
where $\cdots$ denotes a constant (and other non-singular terms). This gives for $G(s)$ the generalized formula
\begin{equation}
    G(s) \propto s \prod_\ell (s + \lambda_\ell).
     \label{eq:gProd}
\end{equation}

Plugging the result from equation \eqref{eq:gProd} into the second term in equation \eqref{eq:bigKeatingResult} and taking appropriate derivatives we find that the rapidly oscillating part of $\disAv{ \rho(E+\omega/2)\rho(E-\omega/2) }_{\text{conn}}$ is
\begin{equation}
    \disAv{ \rho(E+\omega/2)\rho(E-\omega/2) }_{\text{conn}} \supset \frac {\cos 2\pi \omega \hat \rho}{2\pi^2\omega^2}\prod_\ell \frac{\lambda_\ell^2}{\omega^2 +\lambda_\ell^2}.
\end{equation}
This is exactly the non-trivial late-time modification in \eqref{eq:KRMT_F}, recalling that $K(\omega) = 2 \pi \disAv{ \rho(E+\omega/2)\rho(E-\omega/2) }_{\text{conn}}$ in keeping with our Fourier transform conventions.

\begin{figure}
    \centering
    \includegraphics[scale=0.5]{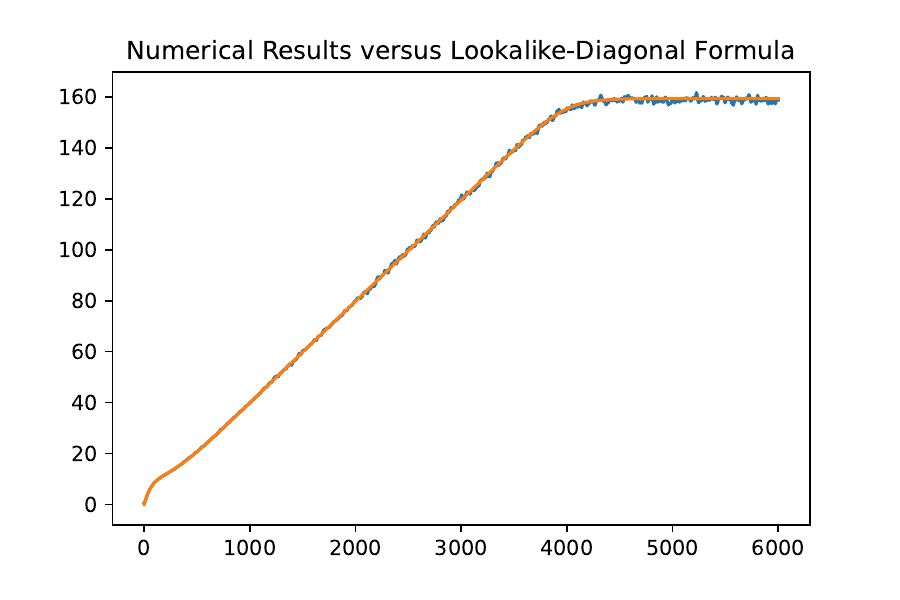}
    \includegraphics[scale=0.5]{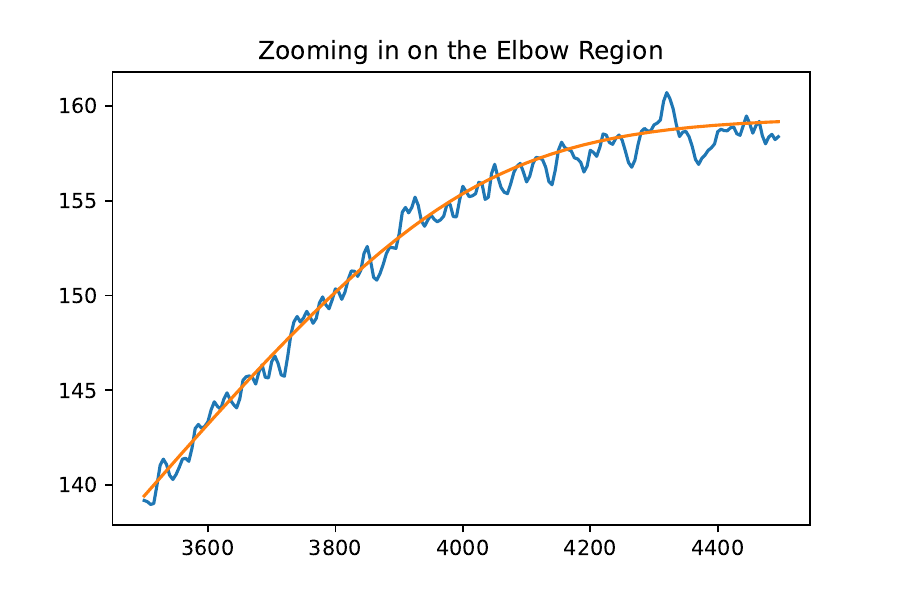}
    \caption{Equation \ref{eq:fullSFF} versus numerical results obtained from 400,000 samples from a random matrix ensemble with four blocks transferring between each other at different rates. The rates were chosen so that the $\lambda$s were degenerate: .0072, .0108, and .0108. The left graph shows the SFF over a wide region, whereas the right graph zooms in on the smoothed out elbow.}
    \label{fig:numericalConfirmation}
\end{figure}

Combined with the earlier time physics of the ramp enhancement, we have all the pieces in \eqref{eq:KRMT_F} and its Fourier transform, \eqref{eq:fullSFF}. One can of course also derive the full function directly using \eqref{eq:bigKeatingResult} and  \eqref{eq:gProd}. We already showed one example of the agreement between theory and numerics in Figure~\ref{fig:IntroPrediction}. Figure \ref{fig:numericalConfirmation} shows equation \eqref{eq:fullSFF} in a more complicated scenarios with four sectors, again with good agreement. 

\section{Outlook}
%%%
\label{sec:outlook}

Ensembles of ``chaotic'' quantum systems are widely expected to have random-matrix-like level repulsion at sufficiently small energy scales. Sometimes the presence of random matrix universality is even taken as a definition of quantum chaos in the general case where no semi-classical limit is available. On the other hand, such physical systems inevitably have structure that causes deviations from pure random matrix theory. One principle source of deviation is the existence of slow modes which cause enhancements in the early-time ramp portion of the SFF. Because of the sum rule, these enhancements must be paid back at later time. For systems in the GUE symmetry class, we put forward a formula, \eqref{eq:fullSFF}, for the SFF at all times which satisfies the sum rule. We also showed that this formula agrees with numerical calculations and can be obtained from the periodic orbit theory for semi-classical quantum chaotic systems. 

There are many applications of the formula presented here. One of the key motivations for our study was the observation in \cite{RichardGlass} of the unexpected suppression of the SFF around the Heisenberg time. Although that work used different methods from the approach pursued here, we have checked in the simplest case of two sectors that our formula is in agreement with the formulas in \cite{RichardGlass}. It would also be interesting to compare our results with other known deviations from RMT, such as those observed in \cite{Jia:2019orl} arising from scale fluctuations in the overall size of the spectrum. 

Another important case we haven't touched on yet is many-body hydrodynamic systems. These systems have one or more local conserved quantities, such as energy or charge. Since the charge in any large subregion is approximately conserved, sectors of the Hilbert space are labelled by an entire smooth density profile. This means that there are exponentially many sectors and an exponentially large enhancement of the ramp at early times. For instance, in diffusive hydrodynamics the enhancement is~\cite{WinerHydro} 
\begin{equation} \trp(T) = \exp \left(V\left(\frac{1}{4\pi DT}\right)^{d/2} \zeta(1+d/2)\right).
\label{eq:diffFactor}
\end{equation}
This exponential factor implies exponentially many slow modes convolved together in equation \eqref{eq:fullSFF}. This, in turn, yields an exponentially wide Gaussian transition from ramp to plateau around the Heisenberg time. It also means that the value of the SFF at the Heisenberg time is exponentially lower than $2\pi \hat \rho$.

Although we understand these qualitative features, the quantitative details are less certain. Presumably the area under the ramp enhancement goes as $e^{cN}$, where $N$ is the system size and $c$ is some non-universal feature of the microscopic theory. For Floquet hydrodynamic systems (discussed more in appendix \ref{app:floquet}) with period $P$, the total area under the SFF enhancement is roughly equal to \eqref{eq:diffFactor} at time $P$. More generally, the exact nature of this UV cutoff time is unclear. Intuitively, the issue is that the early time part of the enhanced ramp is subject to all the non-hydrodynamic short-time physics in the system. It would interesting if there were some way to isolate the hydrodynamic physics of interest when it reappears around the Heisenberg time.

We close with a few comments about potential generalizations and connections. We conjectured that our main result is completely general within the GUE symmetry class provided there is a clean separation between the Thouless time and the Heisenberg time. Hence, it is natural to ask if a more general derivation is possible. The diagonal approximation is a fairly general notion, but the lookalike formula is harder to generalize. Perhaps by merging the effective theory of~\cite{altland2020late} with the hydrodynamic theory of the early time ramp, one could obtain a general formulation. 

Given the fruitful role of spectral statistics in the recent study of black holes and low-dimensional gravity (e.g.~\cite{Saad:2019lba}), it is also interesting to ask what would be gravitational avatars of the late-time reappearance of hydrodynamics. After all, black holes are thermal and hydrodynamic entities (e.g.~\cite{Bhattacharyya:2007vjd}), so they should exhibit an enhanced ramp and hence a subsequent suppression in their SFF (if they could live long enough!).

We also had to restrict to GUE-type systems to obtain our formula, but the sum rule is valid more generally. It would be quite interesting to extend our analysis to GOE-type systems and beyond.

\textit{Acknowledgements:} We thank Amit Vikram, Richard Barney, Victor Galitski, and Chris Baldwin for collaboration on related topics. We thank Steve Shenker, Zhenbin Yang, Shunyu Yao, and Phil Saad for very useful discussions. We acknowledge support from the Joint Quantum Institute (M.W.) and the Air Force Office of Scientific Research under award number FA9550-19-1-0360 (B.S.)

%%%
\appendix
%%%

%%%
\section{Some Fourier transforms}
%%%
\label{app:fourier}

Given a function $F(t)$ in the time domain, we write its Fourier decomposition as
\begin{equation}
    F(t) = \int \frac{d\omega}{2\pi} e^{-i \omega t} F(\omega),
\end{equation}
with the inverse relation being
\begin{equation}
    F(\omega) = \int dt e^{i \omega t} F(t).
\end{equation}

In the time domain, the pure RMT connected energy 2-point function $K_{\text{GUE}}$ in the GUE ensemble with density of states $\hat{\rho}$ is 
\begin{equation}
    K_{\text{GUE}}(t) = \hat{\rho} + \frac{1}{2\pi} \left( |t| - \frac{1}{2}|t+2\pi \hat{\rho}| - \frac{1}{2}|t-2\pi \hat{\rho}| \right).
\end{equation}
The corresponding frequency domain function is
\begin{equation}
    K_{\text{GUE}}(\omega)=2\pi \hat{\rho}\delta(\omega) - \frac{1 - \cos(\omega 2\pi \hat{\rho})}{\pi \omega^2}
\end{equation}

It is also useful to separate out the ramp contribution,
\begin{equation}
    K_{\text{ramp}}(t)=\frac{|t|}{2\pi},
\end{equation}
and
\begin{equation}
    K_{\text{ramp}}(\omega)=-\frac{1}{\pi \omega^2}.
\end{equation}
The latter form is singular at $\omega=0$, so we regulate it as
\begin{equation}
    K_{\text{ramp,reg}}=-\frac{1}{2\pi}\left( \frac{1}{(\omega+i \delta)^2} + \frac{1}{(\omega-i \delta )^2} \right)
\end{equation}
for some positive infinitesimal $\delta$.

When the total return probability (TRP) is significantly different from one, then the ramp is enhanced. For concreteness, we consider a class of systems where the TRP is the exponential of a stochastic matrix, $M$,
\begin{equation}
    \trp(t) = \tr(e^{M |t|}).
\end{equation}
$M$ has one zero eigenvalue and some other number of negative eigenvalues $-\lambda_\ell$. The enhanced ramp is therefore
\begin{equation}
    K_{\text{ramp,enh}}(t)=\frac{|t|}{2\pi}\left(1 + \sum_\ell e^{-\lambda_\ell |t|} \right).
\end{equation}
In the frequency domain we have
\begin{equation}
    K_{\text{ramp,enh}}(\omega)=-\frac{1}{2\pi}\left( \frac{1}{(\omega+i \delta)^2} + \frac{1}{(\omega-i \delta )^2} \right) - \sum_\ell \frac{1}{2\pi}\left( \frac{1}{(\omega+i \lambda_\ell)^2} + \frac{1}{(\omega-i \lambda_\ell )^2} \right).
\end{equation}

%%%
\section{The Two-Point Function from the Generating Function}
%%%
\label{app:Qderiv}

In this section, we give a quick derivation of equation \eqref{eq:rho2Point}, reproduced below:
\begin{equation}
    \disAv{\rho(E+\omega/2)\rho(E-\omega/2) }_{\text{conn}}=-\frac{1}{2\pi^2}\left[\textrm{Re} \frac{\partial^2}{\partial \alpha \partial \beta}Q \right]_{\alpha,\beta , \gamma,\delta = \omega/2}-\frac 12 \bar \rho^2.
\end{equation}
When differentiating with respect to $\alpha$ or $\beta$, one is effectively differentiating the function $\Delta(E)/B(E)$. Since $B$ is purely real and $\Delta(E)$ is defined to still have zeros at the eigenvalues of $H$, it follows that
\begin{equation}
    \textrm{Im} \left[ \frac{ d \log \Delta(E+i\epsilon )}{dE} - \frac{ d \log B(E)}{dE} \right] = \textrm{Im} \left[ \tr \frac{1}{E + i\epsilon -H} \right].
\end{equation}
The latter quantity is the imaginary part of the resolvent; we do not need the real part here (and it anyways typically requires regularization).

Hence, we define
\begin{equation}
    \tilde{R}(E) = \frac{ d \log \Delta(E+i\epsilon )}{dE} - \frac{ d \log B(E)}{dE}
\end{equation}
and immediately have
\begin{equation}
    \textrm{Im } \tilde{R}(E+i\epsilon) = - \pi \rho(E).
\end{equation}
Since $\disAv{\rt}$ is related to $Q$ via
\begin{equation}
    \disAv{\rt}=\partial_s Q(0,s,0,0) |_{s=0}
    \label{eq:rtQ}
\end{equation}
we can plug equation \eqref{eq:bigKeatingResult} into \eqref{eq:rtQ} to get 
\begin{equation}
    \disAv{\textrm{Re }\rt (E+i\epsilon)}=\disAv{\textrm{Re }\rt (E-i\epsilon)}=0.
\end{equation}
This means the real part of $\rt$ is zero on average, while the imaginary part is the density of states.

Let $\delta \rt(E+i\epsilon) = \rt(E+i\epsilon) - ( - i \pi \bar \rho)$ be the fluctuation in $\rt$. We know $\delta \rt(E)$ is analytic in the upper half plane. We also know that the fluctuations decay when $E$ and $E'$ move away from the real axis. Hence, it must be that $\delta \rt(E+i\epsilon)$ is a sum of Fourier waves $A_t e^{i \omega t}$ with $t>0$. The amplitude $A_t$ has translation invariant fluctuations, which requires the real and imaginary parts to have the same variance and to be uncorrelated. Then we have
\begin{equation}
    \disAv{ \delta \rt(E+i\epsilon) \delta \rt(E'+i\epsilon)} =  \int dt  \disAv{ (\textrm{Re }A_t )^2 - (\textrm{Im }A_t)^2 } e^{i t (E+E')} = 0.
\end{equation}
We finally conclude that 
\begin{equation}
\begin{split}
    \disAv{\rho(E+\omega/2)\rho(E-\omega/2) }_{\text{conn}}=\frac{1}{\pi^2} \disAv{\textrm{Im }\delta \rt(E-\omega/2+i\epsilon)\textrm{ Im }\delta \rt (E+\omega/2+i\epsilon)}_{\text{conn}}\\
     = \frac{1}{2\pi^2}\disAv{\delta \rt(E-\omega/2+i\epsilon) \delta\rt (E+\omega/2-i\epsilon)}_{\text{conn}}=-\frac{1}{2\pi^2}\left[\textrm{Re} \frac{\partial^2}{\partial \alpha \partial \beta}Q \right]_{\alpha,\beta , \gamma,\delta = \omega/2}-\frac 12 \bar \rho^2.
\end{split}
\end{equation}

%%%
\section{SFF Suppression In Floquet Circuits}
%%%
\label{app:floquet}

All of the effects in this paper have an analogue for GUE-like Floquet circuits. Now the relevant random matrix ensemble is one of unitary matrices known as the circular unitary ensemble (CUE). The study of spectral statistics of Floquet circuits in the CUE symmetry class has a venerable history including \cite{PhysRevLett.121.060601,Chan:2017kzq,Friedman_2019,moudgalya2020spectral} and \cite{Braun_2012}, the latter of which specifically considers them in a lookalike context. The setup is that we have an $N$-dimensional Hilbert space partitioned into sectors and an ensemble of near-block-diagonal unitaries $U$ with large amplitudes to move within a sector and small amplitudes to move between sectors. The connected SFF is defined for integer $T$ as 
\begin{equation}
    \SFF_{\text{conn}}(T)=\disAv{\tr U^T\tr U^{-T}}_{\text{conn}}.
\end{equation}
The CUE result is 
\begin{equation}
    \SFF_{\text{conn}}(T)= |T|-\textrm{max}(0,|T|-N).
    \label{eq:CUESFF}
\end{equation}
This function has a linear ramp until the Heisenberg time of $N$.

\begin{figure}
    \centering
    \includegraphics[scale=0.4]{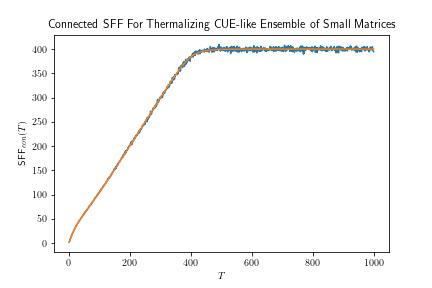}
    \includegraphics[scale=0.4]{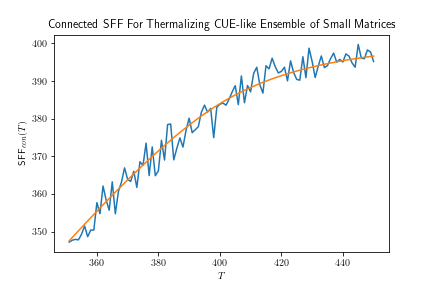}
    \includegraphics[scale=0.4]{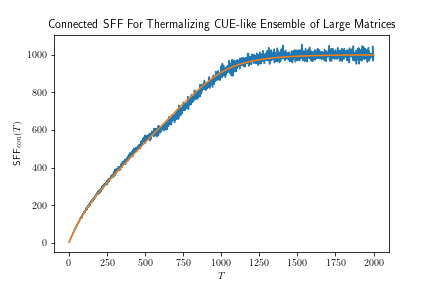}
    \includegraphics[scale=0.4]{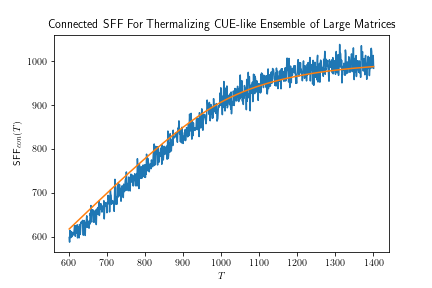}
    \caption{Top Left: The full SFF of an ensemble with two blocks of size $200 \times 200$ and a $\lambda$ of $.031$. $10000$ samples were taken. Top Right: Zooming in on the crossover to the plateau region for the ensemble in Top Left. Bottom Left: The full SFF of an ensemble with two blocks of size $500 \times 500$ and a $\lambda$ of $.005$. $3000$ samples were taken. Bottom Right: Zooming in on the crossover to the plateau region for the ensemble in Bottom Left. Note in particular the kink around $T=500$ in the bottom left graph. This kink occurs because the Thouless time ($200$) is comparable to the single-block Heisenberg time ($500$). This case is not covered in this paper, but it is discussed to some extent in \cite{RichardGlass}.}
    \label{fig:floquetFig}
\end{figure}

Similar to Hamiltonian systems, at short times the SFF for slowly thermalizing Floquet systems is well approximated by~\cite{WinerHydro}
\begin{equation}
    \SFF_{\text{conn}}(T)\approx\trp(T) |T|.
\end{equation}
Modifying the derivation of \eqref{eq:CUESFF} in \cite{Braun_2012} to take slow modes into account gives
\begin{equation}
    \SFF_{\text{conn}}(T)\approx\trp(T) |T|-\frac{(1-e^{-\lambda_1})e^{-\lambda_1 |T|}}{1+e^{-\lambda_1}}*\frac{(1-e^{-\lambda_2})e^{-\lambda_2 |T|}}{1+e^{-\lambda_2}}...*\textrm{max}(0,|T|-N), 
    \label{eq:FloquetResult}
\end{equation}
where now $*$ denotes a discrete convolution. One can show that equation \eqref{eq:FloquetResult} obeys a Floquet analogue of the sum rule for any choice of $\lambda$s. In Figure \ref{fig:floquetFig} we show that equation \eqref{eq:FloquetResult} agrees with numerical results.

\bibliographystyle{ieeetr}
\bibliography{zeta.bib}

\begin{thebibliography}{10}

\bibitem{haake2010quantum}
F.~Haake, {\em Quantum Signatures of Chaos}.
\newblock Springer Series in Synergetics, Springer Berlin Heidelberg, 2010.

\bibitem{PhysRevLett.52.1}
O.~Bohigas, M.~J. Giannoni, and C.~Schmit, ``Characterization of chaotic
  quantum spectra and universality of level fluctuation laws,'' {\em Phys. Rev.
  Lett.}, vol.~52, pp.~1--4, Jan 1984.

\bibitem{mehta2004random}
M.~Mehta, {\em Random Matrices}.
\newblock ISSN, Elsevier Science, 2004.

\bibitem{bohigas1984chaotic}
O.~Bohigas and M.-J. Giannoni, ``Chaotic motion and random matrix theories,''
  in {\em Mathematical and computational methods in nuclear physics},
  pp.~1--99, Springer, 1984.

\bibitem{dubertrand2016spectral}
R.~Dubertrand and S.~M{\"u}ller, ``Spectral statistics of chaotic many-body
  systems,'' {\em New Journal of Physics}, vol.~18, no.~3, p.~033009, 2016.

\bibitem{PhysRevLett.121.264101}
B.~Bertini, P.~Kos, and T.~c.~v. Prosen, ``Exact spectral form factor in a
  minimal model of many-body quantum chaos,'' {\em Phys. Rev. Lett.}, vol.~121,
  p.~264101, Dec 2018.

\bibitem{Chan:2017kzq}
A.~Chan, A.~De~Luca, and J.~T. Chalker, ``{Solution of a minimal model for
  many-body quantum chaos},'' {\em Phys. Rev. X}, vol.~8, no.~4, p.~041019,
  2018.

\bibitem{PhysRevX.8.021062}
P.~Kos, M.~Ljubotina, and T.~c.~v. Prosen, ``Many-body quantum chaos: Analytic
  connection to random matrix theory,'' {\em Phys. Rev. X}, vol.~8, p.~021062,
  Jun 2018.

\bibitem{doi:10.1063/1.1703775}
F.~J. Dyson, ``Statistical theory of the energy levels of complex systems.
  iii,'' {\em Journal of Mathematical Physics}, vol.~3, no.~1, pp.~166--175,
  1962.

\bibitem{wigner1959group}
E.~Wigner and J.~Griffin, {\em Group Theory and Its Application to the Quantum
  Mechanics of Atomic Spectra}.
\newblock Pure and applied Physics, Academic Press, 1959.

\bibitem{PhysRevLett.126.121602}
M.~Srdin\ifmmode~\check{s}\else \v{s}\fi{}ek, T.~c.~v. Prosen, and
  S.~Sotiriadis, ``Signatures of chaos in nonintegrable models of quantum field
  theories,'' {\em Phys. Rev. Lett.}, vol.~126, p.~121602, Mar 2021.

\bibitem{Delacretaz:2022ojg}
L.~V. Delacretaz, A.~L. Fitzpatrick, E.~Katz, and M.~T. Walters,
  ``{Thermalization and chaos in a 1+1d QFT},'' {\em JHEP}, vol.~02, p.~045,
  2023.

\bibitem{Cotler2017}
J.~S. Cotler, G.~Gur-Ari, M.~Hanada, J.~Polchinski, P.~Saad, S.~H. Shenker,
  D.~Stanford, A.~Streicher, and M.~Tezuka, ``Black holes and random
  matrices,'' {\em Journal of High Energy Physics}, vol.~2017, May 2017.

\bibitem{Saad:2018bqo}
P.~Saad, S.~H. Shenker, and D.~Stanford, ``{A semiclassical ramp in SYK and in
  gravity},'' 6 2018.

\bibitem{berry1977level}
M.~V. Berry and M.~Tabor, ``Level clustering in the regular spectrum,'' {\em
  Proceedings of the Royal Society of London. A. Mathematical and Physical
  Sciences}, vol.~356, no.~1686, pp.~375--394, 1977.

\bibitem{KunzShapiro}
H.~Kunz and B.~Shapiro, ``Transition from {Poisson} to {Gaussian} unitary
  statistics: The two-point correlation function,'' {\em Phys. Rev. E},
  vol.~58, pp.~400--406, 1998.

\bibitem{PhysRevLett.125.250601}
Y.~Liao, A.~Vikram, and V.~Galitski, ``Many-body level statistics of
  single-particle quantum chaos,'' {\em Phys. Rev. Lett.}, vol.~125, p.~250601,
  Dec 2020.

\bibitem{RichardGlass}
R.~Barney, M.~Winer, C.~L. Baldwin, B.~Swingle, and V.~Galitski, ``Spectral
  statistics of a minimal quantum glass model,'' 2023.

\bibitem{Winer_Glass}
M.~Winer, R.~Barney, C.~L. Baldwin, V.~Galitski, and B.~Swingle, ``Spectral
  form factor of a quantum spin glass,'' {\em Journal of High Energy Physics},
  vol.~2022, sep 2022.

\bibitem{WinerHydro}
M.~Winer and B.~Swingle, ``{Hydrodynamic Theory of the Connected Spectral form
  Factor},'' {\em Phys. Rev. X}, vol.~12, no.~2, p.~021009, 2022.

\bibitem{bipartite}
F.~Fritzsch and M.~F.~I. Kieler, ``Universal spectral correlations in bipartite
  chaotic quantum systems,'' 2023.

\bibitem{Friedman_2019}
A.~J. Friedman, A.~Chan, A.~De~Luca, and J.~Chalker, ``Spectral statistics and
  many-body quantum chaos with conserved charge,'' {\em Physical Review
  Letters}, vol.~123, Nov 2019.

\bibitem{moudgalya2020spectral}
S.~Moudgalya, A.~Prem, D.~A. Huse, and A.~Chan, ``Spectral statistics in
  constrained many-body quantum chaotic systems,'' 2020.

\bibitem{Gharibyan_2018}
H.~Gharibyan, M.~Hanada, S.~H. Shenker, and M.~Tezuka, ``Onset of random matrix
  behavior in scrambling systems,'' {\em Journal of High Energy Physics},
  vol.~2018, Jul 2018.

\bibitem{PhysRevLett.121.060601}
A.~Chan, A.~De~Luca, and J.~T. Chalker, ``Spectral statistics in spatially
  extended chaotic quantum many-body systems,'' {\em Phys. Rev. Lett.},
  vol.~121, p.~060601, Aug 2018.

\bibitem{Gutzwiller:1971fy}
M.~C. Gutzwiller, ``{Periodic orbits and classical quantization conditions},''
  {\em J. Math. Phys.}, vol.~12, pp.~343--358, 1971.

\bibitem{Berry1981}
M.~Berry, ``Quantizing a classically ergodic system: Sinai's billiard and the
  kkr method,'' {\em Ann. of Phys.}, vol.~131, pp.~163--216, 1981.

\bibitem{MVBerry_1990}
M.~V. Berry and J.~P. Keating, ``A rule for quantizing chaos?,'' {\em Journal
  of Physics A: Mathematical and General}, vol.~23, p.~4839, nov 1990.

\bibitem{10.2307/52022}
J.~P. Keating, ``Periodic orbit resummation and the quantization of chaos,''
  {\em Proceedings: Mathematical and Physical Sciences}, vol.~436, no.~1896,
  pp.~99--108, 1992.

\bibitem{Keating_2007}
J.~P. Keating and S.~Müller, ``Resummation and the semiclassical theory of
  spectral statistics,'' {\em Proceedings of the Royal Society A: Mathematical,
  Physical and Engineering Sciences}, vol.~463, pp.~3241--3250, sep 2007.

\bibitem{PhysRevLett.89.206801}
K.~Richter and M.~Sieber, ``Semiclassical theory of chaotic quantum
  transport,'' {\em Phys. Rev. Lett.}, vol.~89, p.~206801, Oct 2002.

\bibitem{2009NJPh...11j3025M}
S.~{M{\"u}ller}, S.~{Heusler}, A.~{Altland}, P.~{Braun}, and F.~{Haake},
  ``{Periodic-orbit theory of universal level correlations in quantum chaos},''
  {\em New Journal of Physics}, vol.~11, p.~103025, Oct. 2009.

\bibitem{2005PhRvE..72d6207M}
S.~{M{\"u}ller}, S.~{Heusler}, P.~{Braun}, F.~{Haake}, and A.~{Altland},
  ``{Periodic-orbit theory of universality in quantum chaos},'' {\em pre},
  vol.~72, p.~046207, Oct. 2005.

\bibitem{Braun_2012}
P.~Braun and F.~Haake, ``Chaotic maps and flows: exact
  riemann{\textendash}siegel lookalike for spectral fluctuations,'' {\em
  Journal of Physics A: Mathematical and Theoretical}, vol.~45, p.~425101, oct
  2012.

\bibitem{10.2307/52059}
M.~V. Berry and J.~P. Keating, ``A new asymptotic representation for zeta
  (1/2+it) and quantum spectral determinants,'' {\em Proceedings: Mathematical
  and Physical Sciences}, vol.~437, no.~1899, pp.~151--173, 1992.

\bibitem{bogomolny}
E.~B. Bogomolny, ``Semiclassical quantization of multidimensional systems,''
  {\em Nonlinearity}, vol.~5, p.~805, jul 1992.

\bibitem{1994JETPL..60..656K}
V.~E. {Kravtsov} and A.~D. {Mirlin}, ``{Level statistics in a metallic sample:
  corrections to the Wigner-Dyson distribution},'' {\em Soviet Journal of
  Experimental and Theoretical Physics Letters}, vol.~60, p.~656, Nov. 1994.

\bibitem{1995PhRvL..75..902A}
A.~V. {Andreev} and B.~L. {Altshuler}, ``{Spectral Statistics beyond Random
  Matrix Theory},'' {\em prl}, vol.~75, pp.~902--905, July 1995.

\bibitem{2000PhRvL..85.5615A}
A.~{Altland} and A.~{Kamenev}, ``{Wigner-Dyson Statistics from the Keldysh
  {\ensuremath{\sigma}}-Model},'' {\em prl}, vol.~85, pp.~5615--5618, Dec.
  2000.

\bibitem{2020Prosen}
D.~Roy and T.~Prosen, ``Random matrix spectral form factor in kicked
  interacting fermionic chains,'' {\em Physical Review E}, vol.~102, Dec 2020.

\bibitem{Winer_2020}
M.~Winer, S.-K. Jian, and B.~Swingle, ``Exponential ramp in the quadratic
  sachdev-ye-kitaev model,'' {\em Physical Review Letters}, vol.~125, dec 2020.

\bibitem{Roy_2022}
D.~Roy, D.~Mishra, and T.~Prosen, ``Spectral form factor in a minimal bosonic
  model of many-body quantum chaos,'' {\em Physical Review E}, vol.~106, aug
  2022.

\bibitem{halfWH}
P.~Saad, S.~H. Shenker, D.~Stanford, and S.~Yao, ``Wormholes without
  averaging,'' 2021.

\bibitem{Garcia_Garcia_2022}
A.~Garcia-Garcia and V.~Godet, ``Half-wormholes in nearly
  {AdS}{\textdollar}{\_}2{\textdollar} holography,'' {\em {SciPost} Physics},
  vol.~12, apr 2022.

\bibitem{Rosenzweig:1960zz}
N.~Rosenzweig and C.~E. Porter, ``{'Repulsion of Energy Levels' in Complex
  Atomic Spectra},'' {\em Phys. Rev.}, vol.~120, pp.~1698--1714, 1960.

\bibitem{10.1007/3-540-17171-1_1}
M.~V. Berry, ``Riemann's zeta function: A model for quantum chaos?,'' in {\em
  Quantum Chaos and Statistical Nuclear Physics} (T.~H. Seligman and
  H.~Nishioka, eds.), (Berlin, Heidelberg), pp.~1--17, Springer Berlin
  Heidelberg, 1986.

\bibitem{edwards1974riemann}
H.~M. Edwards, {\em Riemann's Zeta Function}.
\newblock Academic Press, 1974.

\bibitem{montgomery2017exploring}
H.~Montgomery, A.~Nikeghbali, and M.~T. Rassias, eds., {\em Exploring the
  Riemann Zeta Function: 190 years from Riemann's Birth}.
\newblock Springer, 2017.

\bibitem{apostol1998introduction}
T.~Apostol, {\em Introduction to Analytic Number Theory}.
\newblock Undergraduate Texts in Mathematics, Springer New York, 1998.

\bibitem{zbMATH02552973}
C.~L. Siegel, ``{\"U}ber {Riemanns} {Nachla{{\ss}}}\ zur analytischen
  {Zahlentheorie}..'' Quellen u. {Studien} {B}. 2, 45-80 (1932)., 1932.

\bibitem{montgomery1973}
H.~L. Montgomery, ``The pair correlation of zeros of the zeta function,'' {\em
  Analytic number theory}, vol.~24, pp.~181--193, 1973.

\bibitem{Goldston1987}
D.~A. Goldston and H.~L. Montgomery, {\em Pair Correlation of Zeros and Primes
  in Short Intervals}, pp.~183--203.
\newblock Boston, MA: Birkh{\"a}user Boston, 1987.

\bibitem{leboeuf2001riemannium}
P.~Leboeuf, A.~G. Monastra, and O.~Bohigas, ``The riemannium,'' 2001.

\bibitem{riemannReview}
M.~V. Berry and J.~P. Keating, ``The riemann zeros and eigenvalue
  asymptotics,'' {\em SIAM Review}, vol.~41, no.~2, pp.~236--266, 1999.

\bibitem{Waltner_2019}
D.~Waltner and K.~Richter, ``Towards a semiclassical understanding of chaotic
  single- and many-particle quantum dynamics at post-heisenberg time scales,''
  {\em Physical Review E}, vol.~100, oct 2019.

\bibitem{Haake}
F.~Haake, {\em Quantum Signatures of Chaos}, vol.~54.
\newblock 01 2001.

\bibitem{Hannay_1984}
J.~H. Hannay and A.~M. O.~D. Almeida, ``Periodic orbits and a correlation
  function for the semiclassical density of states,'' {\em Journal of Physics
  A: Mathematical and General}, vol.~17, p.~3429, dec 1984.

\bibitem{Jia:2019orl}
Y.~Jia and J.~J.~M. Verbaarschot, ``{Spectral Fluctuations in the
  Sachdev-Ye-Kitaev Model},'' {\em JHEP}, vol.~07, p.~193, 2020.

\bibitem{altland2020late}
A.~Altland and J.~Sonner, ``Late time physics of holographic quantum chaos,''
  2020.

\bibitem{Saad:2019lba}
P.~Saad, S.~H. Shenker, and D.~Stanford, ``{JT gravity as a matrix integral},''
  3 2019.

\bibitem{Bhattacharyya:2007vjd}
S.~Bhattacharyya, V.~E. Hubeny, S.~Minwalla, and M.~Rangamani, ``{Nonlinear
  Fluid Dynamics from Gravity},'' {\em JHEP}, vol.~02, p.~045, 2008.

\end{thebibliography}

\end{document}